# Enhanced Capacitance in Surfactant Mediated Ion Complexes

Jihua Chen,[1] Alejandro Espera,[2] Jan Michael Y. Carrillo,[1] Rigoberto Advincula[1,3]

[1]Center for Nanophase Materials Science, Oak Ridge National Laboratory
[2]Engineering Fundamentals, Tickle College of Engineering, the University of Tennessee Knoxville
[3]Department of Chemical and Biomolecular Engineering, University of Tennessee, Knoxville

## Abstract

Ion complexes hold the key for various energy transfer and communication systems in nature and industries. Achieving controllable mechanical and electrical properties in these complex systems is highly desirable but remains challenging. In this work, we propose the use of amphiphilic molecules to mediate salt crystallization. The resultant ionic interfaces can be tunable through reduced surface tensions of crystal facets, an additional intermolecular modifier, and the diffusion-limited crystallization during controlled solvent evaporation. Specifically, we utilized the ionic surfactant Sodium Lauryl sulfoacetate (SLSA) to mediate the crystallization of sodium chloride. Citric acid (CA) was adopted as a hydrogen-bond modifier. Systems mediated by SLSA exhibited the highest capacitance, and a significant enhancement of capacitance was also observed in systems with both SLSA and CA. The substantially increased capacitance of these ionic complexes can be attributed to changes in interfacial and crystalline grain structures. Transmission electron microscopy (TEM), optical microscopy (OM), and Finite Element Analysis (FEA) were used to study the effects of surfactant molecules in these ionic complexes. Understanding the role of ionic complexation in defining the thermodynamic and kinetic process of the crystallization process will contribute to better optimization in nucleation and scalability of organic and inorganic crystal production.

# 1. Introduction

Ionic complexes are critical components  in living systems. The crystallization of ionic crystals, such as calcium carbonate and sodium chloride, in the presence of surfactants or additives, is of intense scientific interest and industrial importance.[1] An ionic complexes with tailorable electrical properties can be important for a variety of energy conversion processes.[2-4] Controlled crystallization is central in the formation of hard structures influencing processes such as biomineralization, bone-growth, mining and flocculation and sedimentation phenomena, drug production, etc.

In this work, we use an ionic surfactant sodium Lauryl sulfoacetate (SLSA) to guide the crystallization of sodium chloride. **Figure** 1 shows the 2D and 3D molecular structure of SLSA.[5] In the presence of surfactant, it is known that the crystallization front will have a reduced surface tension and lead to a preferred growth facet that may not be otherwise achievable.[6-7]  Citric acid (CA) was also added into the system to modify the strength of hydrogen bonding. Up to 10 times higher normalized capacitance was observed in SLSA/ NaCl complexes. This enhanced electrical performance in ionic complexes were attributed to changes in interfacial and crystalline structures. The effects of ionic surfactant in these crystalline ion complexes were examined with Transmission electron microscopy (TEM), optical microscopy (OM), and Finite Element Analysis (FEA).

**Figure** 1. The 2D and 3D molecular structure of sodium Lauryl sulfoacetate, or SLSA.[5]

# 2. Results and Discussion

## 2.1 Increased Capacitance in SLSA/NaCl

The mix ratios of Citric Acid (hydrogen bonding modifier), SLSA, and NaCl are used as their identifier. (**Table** 1) For example, xyz corresponds to a weight ratio of x:y:z in the original total solid content. A stock solution is made with 5-step filtered water, at a concentration of 1%wt total solid content. It is noted that, although Citric Acid/NaCl is also made (xyz=208), the final film degrades in a day. It is not tested and compared in this study against other compositions such as (xyz= 001,  028, and 118).

**Table** 1. Composition ratio of the SLSA/NaCl systems studied in this work. The weight ratio of Citric Acid (hydrogen bonding modifier), SLSA, and NaCl are used as their identifier: x, y, and z, respectively.

| **Name/ Weight ratio** | **x (citric acid)** | **y (SLSA)** | **z (NaCl)** | **Note** |
|---|---|---|---|---|
| 001 | 0 | 0 | 1 | |
| 028 | 0 | 2 | 8 | |
| 208 | 2 | 0 | 8 | Films degrades after formation. |
| 118 | 1 | 1 | 8 | |

Subsequently, the SLSA/NaCl systems are grown on metal plates of 1 cm diameter for electrical testing. A typical procedure for the slow crystallization of SLSA/NaCl is described in the Experimental section. Capacitance is proportional to dielectric constant (*k*), area *A*, and the inverse of distance *d*, through the constant $\varepsilon_0$ of $8.854\times10^{-12}\ \mathrm{F\cdot m^{-1}}$(**Equation** 1).

$$c = k\varepsilon_0\frac{A}{d} \qquad \textbf{Equation } (1)$$

Thus, the normalized capacitance in thin films, or dielectric constant k will be calculated by using **Equation** 2. The term “normalized capacitance” is preferred, because this is a study on thin films with morphological variations. In comparison, dielectric constant is an intrinsic property that is free of the impact from such effects. The same measurement is also done on a few known materials (such as paper of similar thickness) for verification and calibration purposes.

$$k = \frac{c\, d}{\varepsilon_0 A}$$ **Equation** (2)

**Figure** 2 compared the capacitance measurements of the SLSA/NaCl systems. Normalized capacitance is free of thickness effect, equivalent to a dielectric constant value in thin films. SLSA/NaCl (Sample 028), without including any Citric acid in the system, demonstrates an almost 20-fold increase in average values, while maintaining a low standard deviation (5%). Sample 118 (Citric acid:SLSA: NaCl of 1:1:8), on the other hand, also has a normalized capacitance averaging 14, with a much larger standard deviation, due to its morphological variations.

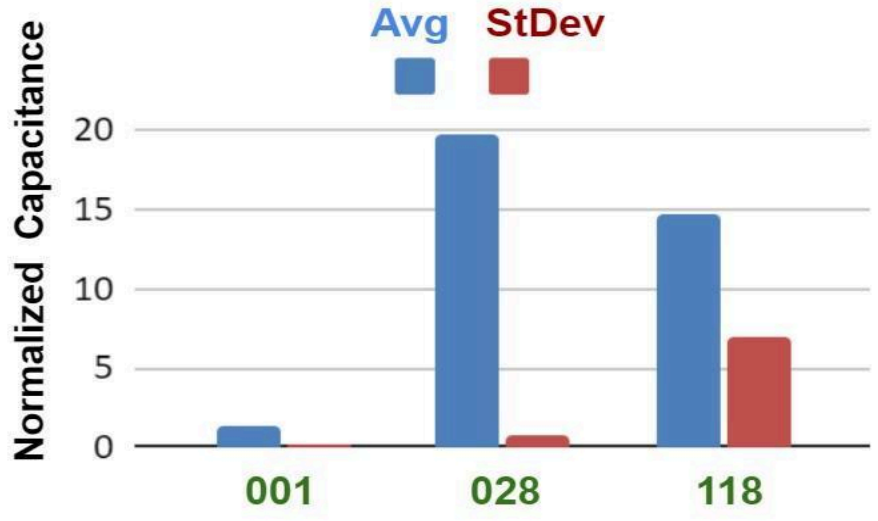

**Figure** 2. Normalized capacitance of different SLSA/NaCl systems. The short name xyz corresponds to a weight ratio of x:y:z in the original total solid content (Citric acid: SLSA: NaCl). It is noted that, although Citric Acid/NaCl is also made (xyz=208), the final film degrades soon afterwards. It is not tested and compared in this study against other compositions such as (xyz= 001, 028, and 118). Five measurements are made for each data point.

## 2.2. Optical Micrographs

Sample 001 (NaCl without additives) has a large morphological variation as shown in **Figure** 3 a and b. Sometimes the crystal is on the order of 100 microns and almost 3D, while in many other cases, their sizes are below 10 microns and there can be a series of Hopper crystals with connected path between a network of much smaller crystals. Sample 208 (citric acid: NaCl of 2:8 by weight ratio) is not compared in Figure 3 due to its strong tendency of degradation in air. Sample 028 (SLSA: NaCl of 2:8 by weight ratio) seems to be more consistent in film morphology, and the resultant NaCl crystallite facets can be much less distinguishable (**Figure** 3c). Sample 118 (citric acid: SLSA: NaCl of 1:1:8), in **Figure** 3d, yields a highly fractal and hierarchical network of crystals, highlighting the combined effect of surface tension reduction during crystallization and hydrogen bond modification through citric acid. The final morphology of ionic crystalline network seem to be highly dependent upon crystallization temperature, crystallization duration, composition and intermolecular interactions of the cocrystallization systems, and substrate surface hydrophobicity, comparable to cocrystallization of molecular crystals. [8-14] This tunability corresponds well to the enhanced capacitance in **Figure** 2. Fractal and hierarchical nature of Sample 118 corresponds to its higher performance variation, while the consistent morphology of Sample 028 contributes to its performance consistency in Figure 2.

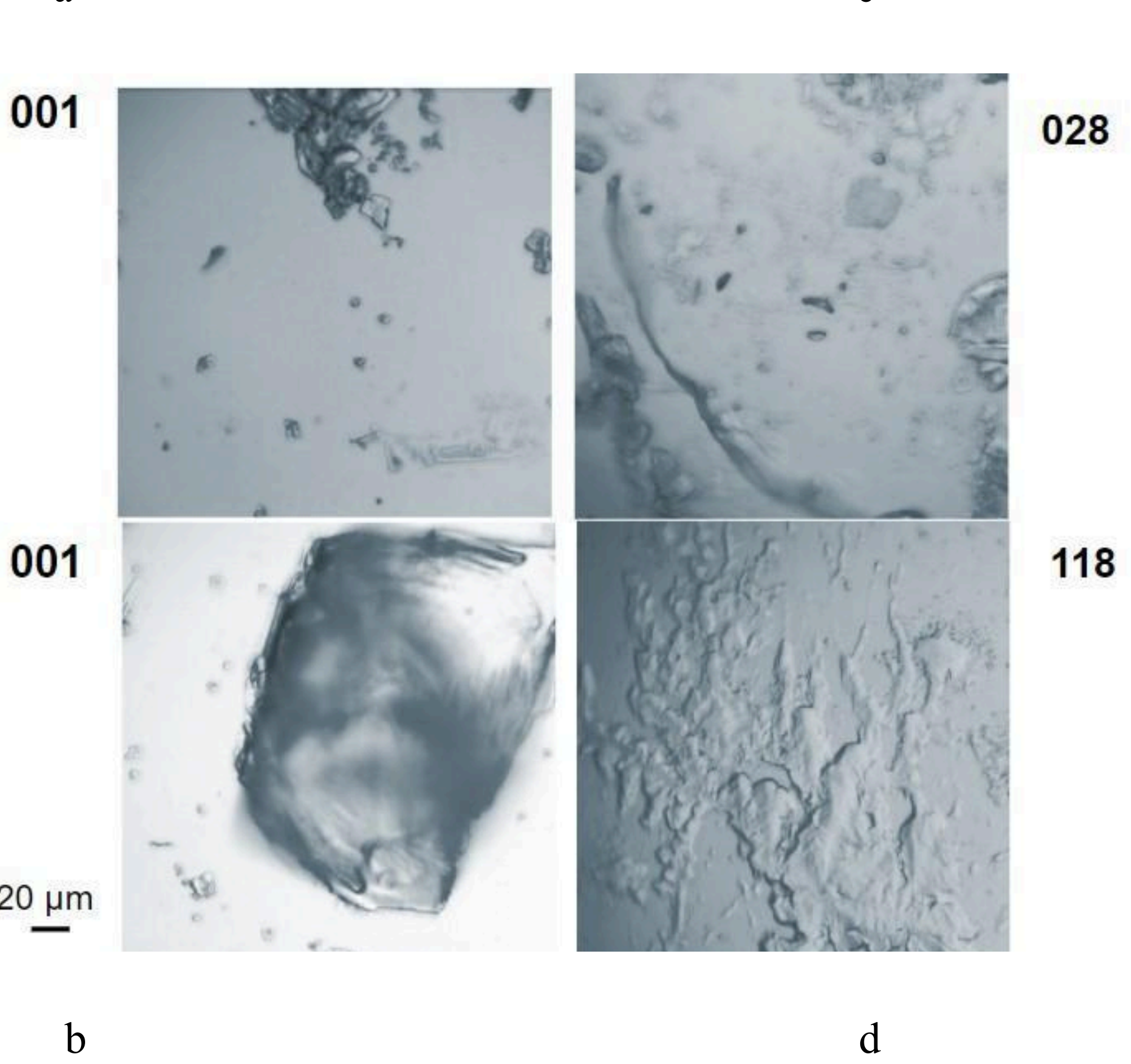

**Figure 3.** Optical micrographs of SLSA/NaCl crystals. The short name xyz corresponds to a weight ratio of x:y:z in the original total solid content (Citric acid: SLSA: NaCl). It is noted that, although Citric Acid/NaCl is also made (xyz=208), the final film degrades soon afterwards. It is not tested and compared in this study against other compositions such as (xyz= 001, 028, and 118).

A typical NaCl crystal without additive has (001) facets, while modified surface tension during additive-mediated crystallization can lead to (110) and (111) facets as shown in **Figure** 4.[7] This leads to needle or wire-shaped NaCl crystals. It is likely that these facets contributed to the fractal and hierarchical microstructures of Sample 118 (**Figure** 3d), impacting on its capacitance and electrical properties.

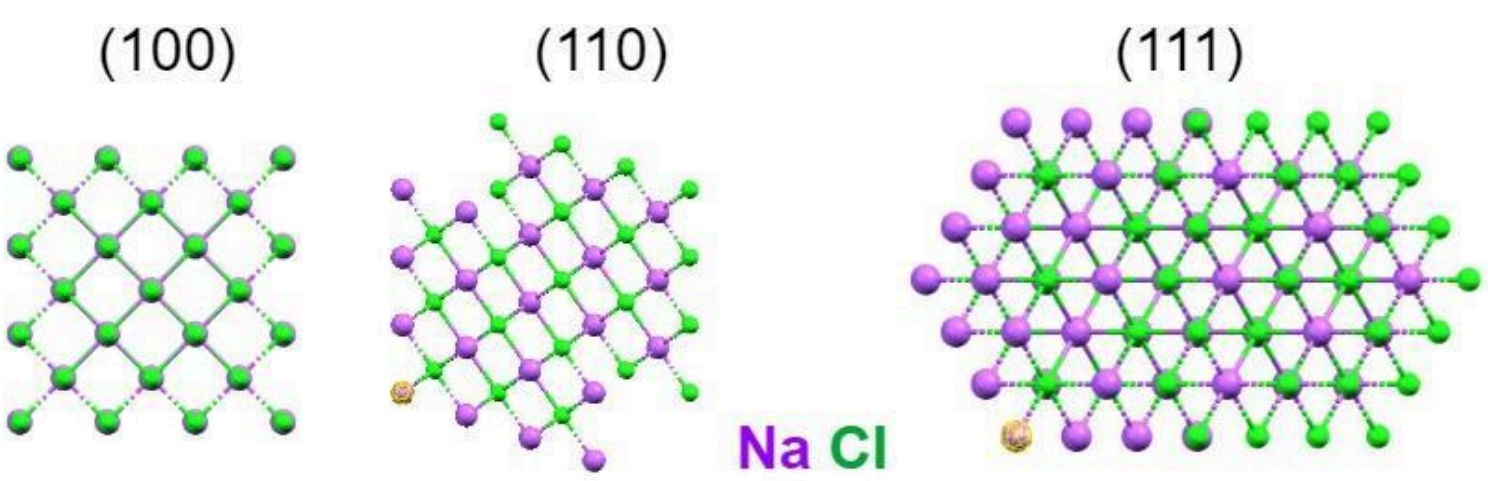

**Figure 4.** Possible facets of NaCl crystals. A typical NaCl crystal without additive has (001) facets, while modified surface tension during additive-mediated crystallization can lead to (110) and (111) facets, which can contribute to the observations shown in **Figure** 3.[7]

## 2.3. TEM and Semantic Segmentation

Although identical slow crystallization may not be possible on TEM grids, a similar concentration and composition ratio is adopted. The final crystal dimensions on TEM grids can be orders of magnitude smaller than the ones on glass slide, but the resultant samples yielded a suitable thickness for TEM, ideal for local structural examination of surfactant mediated salt crystallization. **Figure** 5a shows a TEM image of Sample 028 (SLSA/NaCl). **Figure** 5b has possible locations of the ion complexes highlighted in red. A false color enhanced version is also given in **Figure** 5c. Although the contrast between surfactant and carbon substrate of TEM grid is poor in the original figure, the atomic-number and mass-thickness differences between salt elements and the rest of components are significant. One can easily identify the salt crystal from carbon-based materials, simply because salts allow much less transmitted electrons and give much darker region. **Figure** 5b, on the other hand, segmented possible locations of the ion complexes (red color) by applying a FFT band filter pass, and these correspond to the weakly visible round-shaped surfactant area plus the salt area. The false color enhanced version of the original image yields better boundaries of such regions *c.*(**Figure** 5c).

b. a.

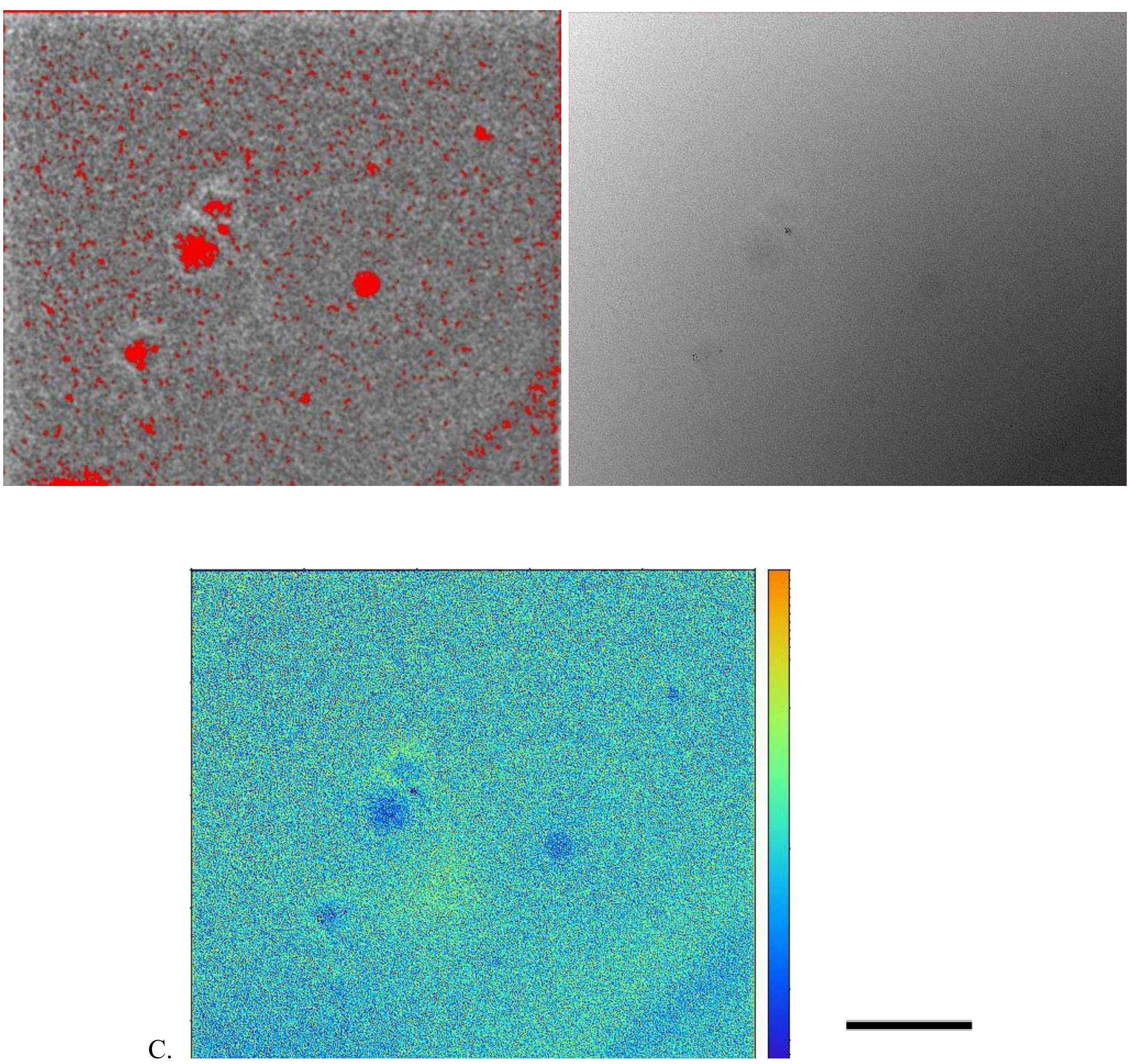

**Figure 5.** TEM image (a) of Sample 028, with segmentation (b) showing possible locations of the ion complexes. A false color enhanced version is also given in **Figure** 5c. The scale bar is 400 nm.

## 2.4. Finite Element Analysis

Controlled crystallization in this work involves two important aspects: fluid movement and heat transfer. The former relates to the balance between convective and diffusive processes, and the later involves a localized temperature gradient.

The fluid movement is driven by the evaporation process during the solution crystallization.[7] Convective movement brings $Na^+$ or $Cl^-$ ions close to the crystallization front, while diffusion matches suitable ions with a lattice position, provided that a minimal surface energy is achieved. A convective-dominant crystallization prefers cubic growth at low supersaturation since there are always fresh supplies of ions near possible lattice sites, while a diffusion-controlled process tends to have planar, or stepped Hopper growth as the ion supplies deplete quickly around growth crystals.[7] **Figure** 6 illustrates the possible effects of surfactant on NaCl crystallization in terms of convective flow. To simplify the comparison, a 2D simulation box of 100 by 100 micron is assumed, along with hypothetical NaCl crystals of cubic shapes and surfactant coverage of round boundaries. Brighter colors indicate a sharper change in fluid movement. The arrow depicted the potential flow direction. It is shown that surfactant can lead to more turbulent changes in convective flow, promoting a diffusion-dominant scenario. This can in turn explain the Hopper-like crystal networks in Sample 118 (**Figure** 2d).

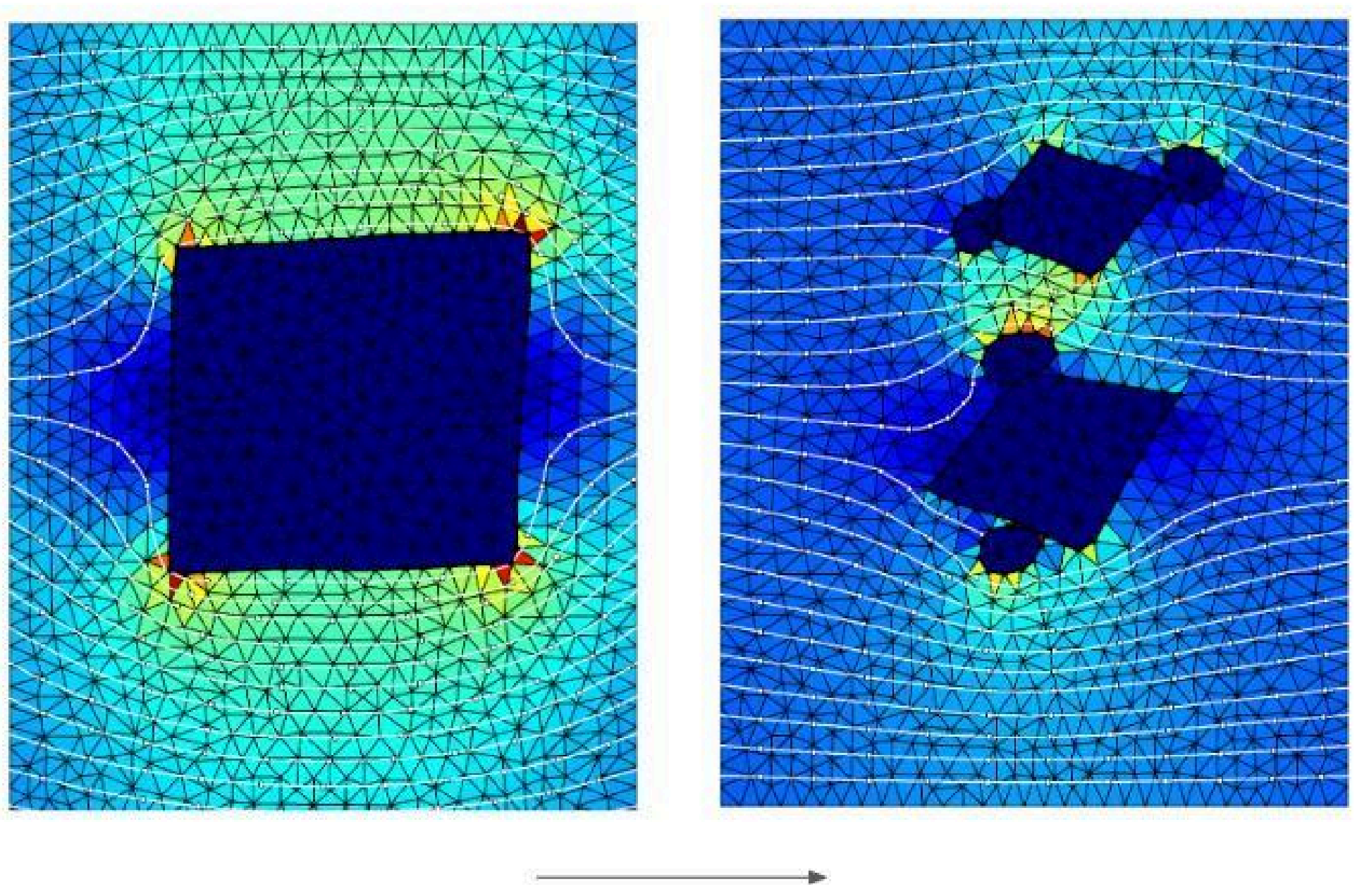

**Figure 6.** FEA results of fluid speed in potential flow. Left: NaCl only, Right: NaCl with surfactant. The arrow depicted the potential flow direction. Brighter colors indicate a sharper change in fluid movement.

Heat maps of FEA show the added surfactants can greatly affect the temperature distribution during the solution crystallization process (**Figure** 7). Brighter color in these heat maps indicates faster heat changes during solution crystallization. This is due to the fact that salt has lower heat capacity than salt/water mixture and pure water.[15-17] The effect of surfactant on temperature gradient and spatial distribution in turn affects nucleation and crystallization, adding extra complexity on the resultant crystalline network and capacitance performance (**Figure** 2 and 3).

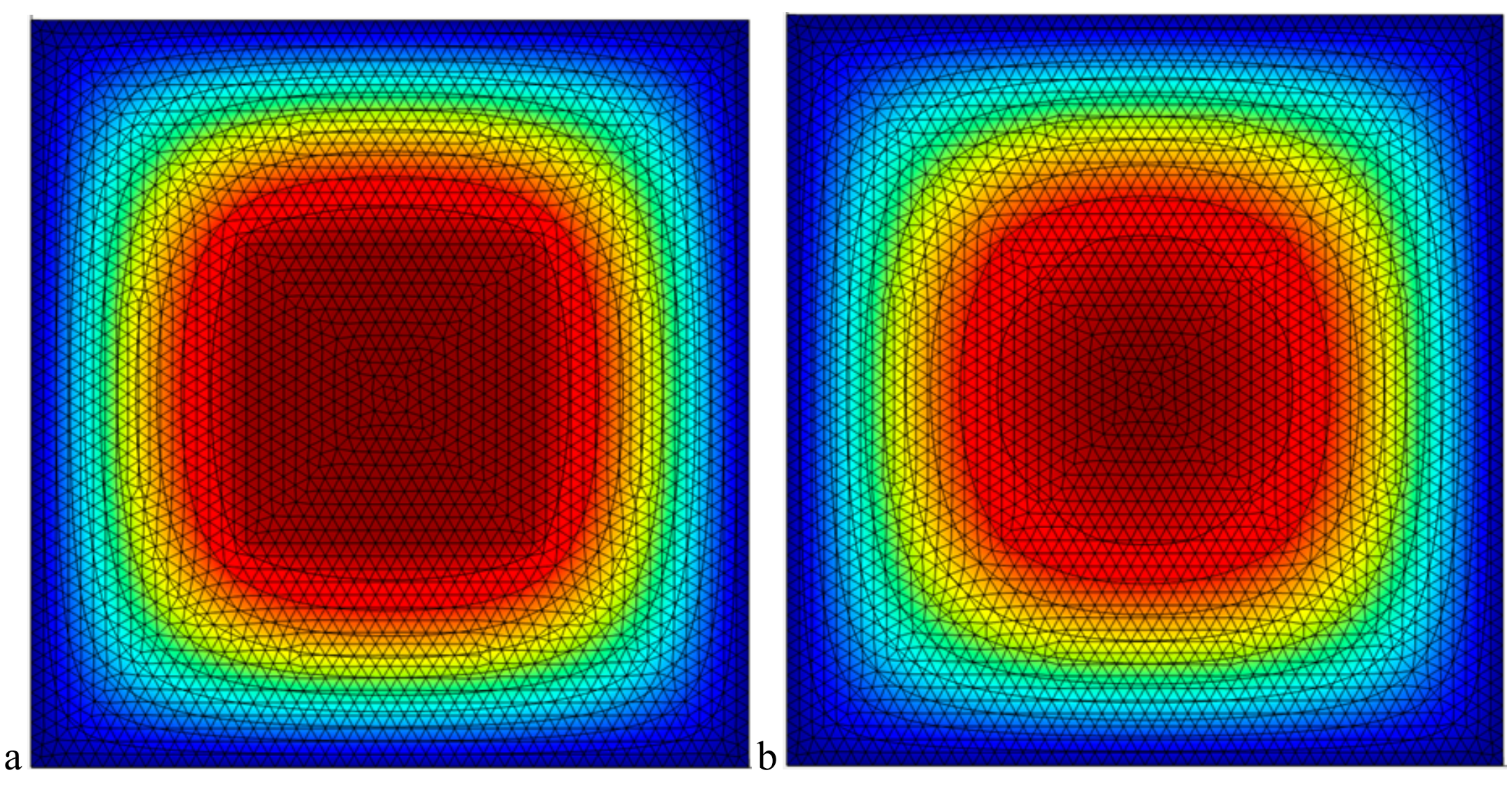

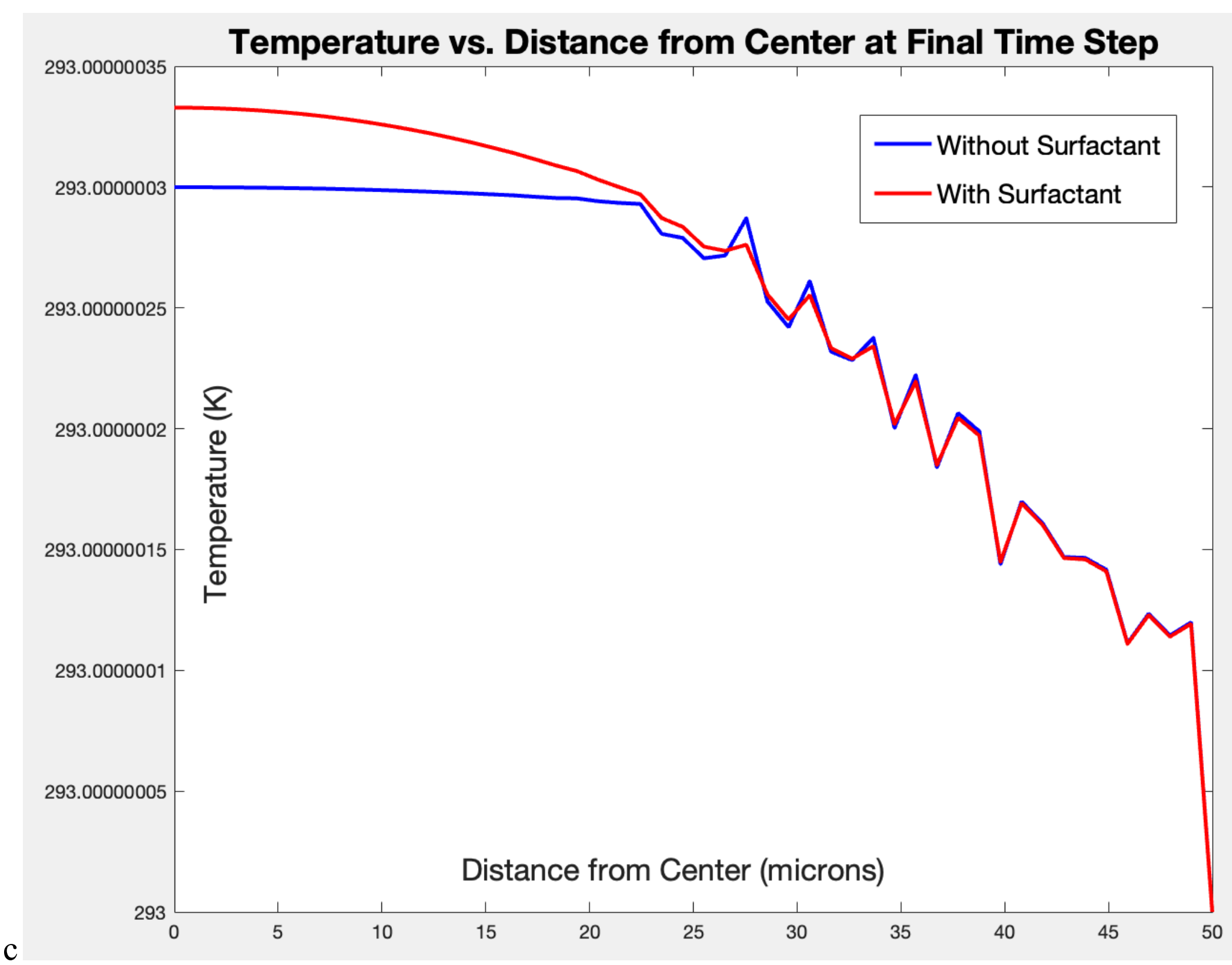

**Figure 7.** FEA results of heat map. a). NaCl only, b). NaCl with surfactant. Brighter color indicates faster heat changes during solution crystallization. c). A plot of temperature T(r) as a function of r, where r is the distance from the crystal center for 2D NaCl crystals with (Red) and without surfactant (Blue). One can clearly see the difference in temperature between with and without surfactant from the center of the crystal up to 20 um (edge of the crystal). At that cut-off distance is the interface of surfactant and water, beyond which would all be water – where no significant difference in temperature can be observed.

## 2.5 Crystal dimensions and nucleation density

The final distribution of crystal dimension and nucleation density (**Figure** 8) is a combined effect from the fluid movement and temperature gradient as discussed in the previous section. The values of crystal dimensions (*D*) and nucleation per unit area (*N*, or the whole area of a micrograph at the same magnification) are extracted from optical micrographs (**Figure** 2). The surfactant and additives in Sample 028 and 118 seem to lower the average value and standard deviation of crystal dimension *D*, generating films of much more uniformity. At the same time, they lead to 4-fold increases in nucleating density. The balance of crystal dimension and nucleation that they achieved can be attributed to the capacitance improvement in **Figure** 2.

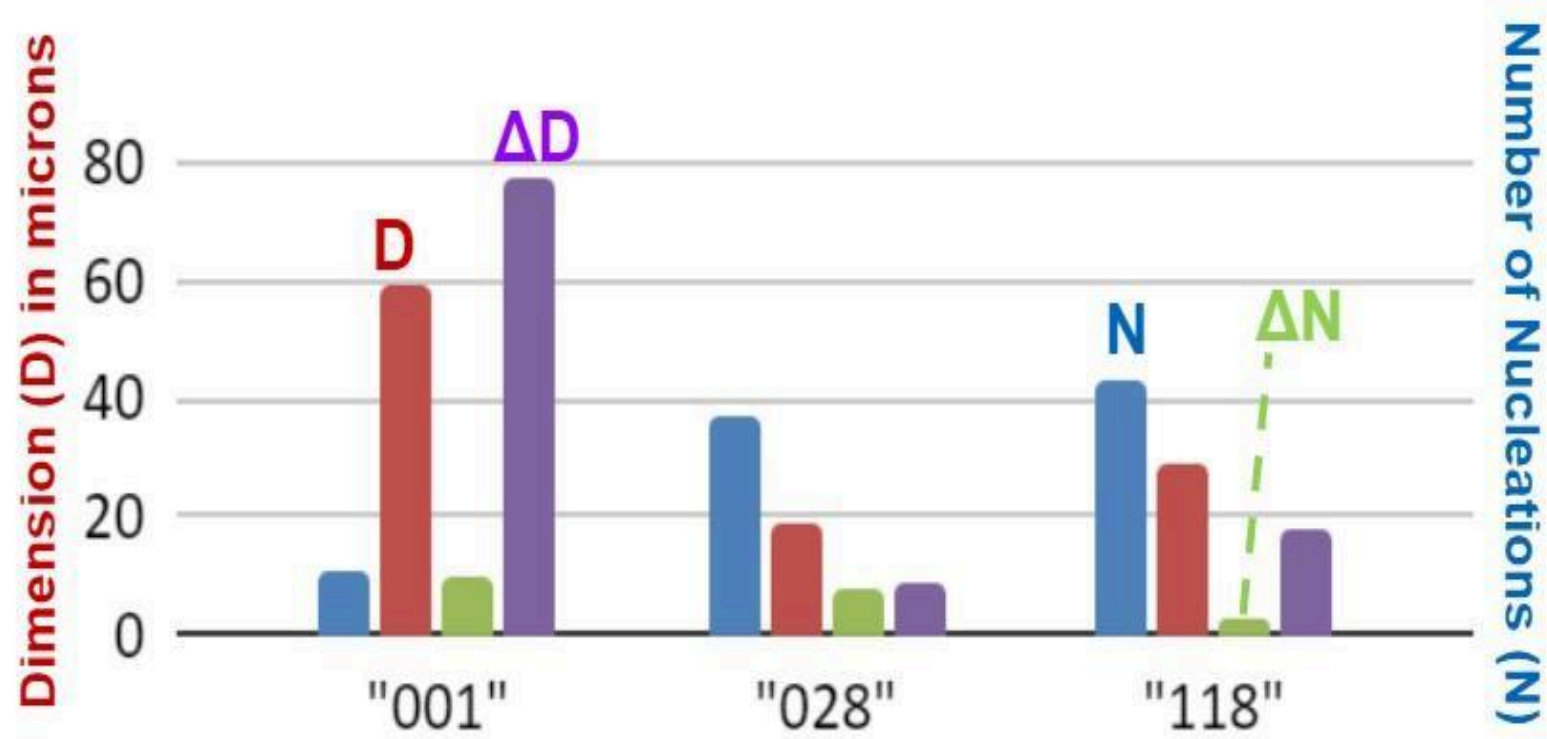

**Figure 8**. Crystal dimensions and nucleation per unit area (i.e. the whole area of a micrograph at the same magnification), extracted from optical micrographs. The short name xyz corresponds to a weight ratio of x:y:z in the original total solid content (Citric acid: SLSA: NaCl). It is noted that, although Citric Acid/NaCl is also made (xyz=208), the final film degrades soon afterwards. It is not tested and compared in this study against other compositions such as (xyz= 001, 028, and 118). About 20 measurements are made for each data point.

## 2.6 Discussions

Other notable electrical characteristics of an ion complex system may involve dielectric spectroscopy, impedance measurements, conductivity, and permittivity.[18] Thin film capacitors are previously reported to have a capacitance range from 10 nF to 0.1mF with various restrictions on performance consistency, mechanical properties, and processing requirements.[19] This work demonstrated a novel way to fabricate organic or ionic complexes as a thin film capacitors, or solid electrolytes for future energy storage systems.

This work used a traditional parallel plate configuration [19] for capacitance measurements. Crystal sizes have significant impacts on the resultant thin film performance [8,11-14] and remain the focus of this work as shown in Figure 2, 3, and 8. Refractive index of the different components of our ternary systems can contribute to the optical contrast that we observed in Figure 3, along with thickness and interfaces. Birefringence or polarized measurements can be very useful for samples with uniform thickness and composition, while our ionic complexes are studied without birefringence to simplify our morphological studies of the ternary blends. Future high resolution TEM or cryogenic TEM may shed light on the chemical nature of the surfactant-salt interfaces, although the electron-beam-sensitive nature of the ion complexes can pose significant challenges for nanoscale chemical imaging. Finite element analysis of salt crystallization is scarce in the literature, and we only find one publication in the context of salt creeping in porous media.[20]

According to a previous work using both ionic and non-ionic surfactants for salt crystallization, surfactants do not act as nucleation sites in solution, but will inhibit nucleation, reduce surface tensions of salt facets and induce skeletal growth.[21] Furthermore, Qazi, *et al.,* [7] reported inclusion of nonionic surfactant molecules, but not ionic ones, in salt crystals, while their sum frequency generation spectroscopy results suggested both ionic and non-ionic surfactant presences at the solid-liquid interface. The adsorption of surfactants in a diffusion-limited salt crystallization process is expected to passivate nucleations near crystal-solution and air-solution interfaces, without significant impacts on solvent evaporation rate. To further understand the nature of the surfactant-salt interfaces, future experimental designs may involve using surface sensitive methods such as Sum Frequency Generation (SFG) and surface plasmon microscopy on a surfactant-grown single crystal.

By varying the specific interactions between the surfactant and salt, as well as an interaction modifier, we expect that further improvements of electrical performances in these salt complexes may be realized and optimized for a wide range of applications such as electrochemical devices, solid electrolytes, and thin film capacitors.

# 3. Conclusion

We show that surfactant-mediated salt crystallization provides a promising way towards electrical property tuning. SLSA/NaCl systems are studied with and without a hydrogen bond modifier-- Citric Acid. SLSA mediated salt crystallization yields up to 10-fold enhancement in normalized capacitance of the resultant thin films. Optical microscopy, TEM, and FEA suggest that a balanced surfactant-salt network is achieved from a combined effect of temperature gradient, diffusive and convective flow. Future optimization is possible for specific types of salt and surfactant in the context of a more specialized application such as solid electrolyte, flexible capacitors, and other ion-related energy applications.

# 4. Experimental

**Materials**

Citric Acid (pharmaceutical grade, Wellbody Naturals, CA), Sodium Lauryl sulfoacetate or SLSA (Pure Organic Ingredients, Utah), and Sodium Chloride (reagent grade, Aldon Corporation, NY),of desired amount are dissolved in 5-stage filtered water to yield the target weight ratio of 0:0:1 (pure salt), 0:2:8 (no citric acid), 2:0:8 (no surfactant), and 1:1:8 (ternary blend). As short names, those sample blend ratios will be referred to as 001, 028, 208, and 118, to mark the weight ratios of citric acid, SLSA surfactant, and salt. The final solutions have a total solid content of 1%wt.

**Controlled Crystallization**

A typical crystallization is performed as the following. A small amount (50 microlitre) of the mixture solution is applied on substrates (such as Amscope BS-50P-100S-22 precleaned glass slide for optical microscopy, or 15mm diameter stainless steel plate of 18 Gauge thickness for capacitance measurements). The substrate is placed in a petri dish with cover, and heated at 50 degree C. A typical crystallization lasts up to a few hours. A baking is performed after the final formation of the resultant film to evaporate off the residual water for 6 hours at the same temperature with the cover removed.

**Capacitance Measurements**

A Proster LCR meter, with a measuring range of pico to milli Faraday,  is used to measure the capacitance of sandwiched metal-sample-metal fixtures. The connection is made through highly conductive stainless steel wire (Master Wire Supply, 316L, 22 Gauge Wire). The Proster LCR meter has a manufacturer certified resolution of 0.1 pF and 2.5% accuracy. Thickness measurements are done with a Beslands Digital Electronic Display Micrometer and typical measured film thickness is on the order of 100 micrometers. Capacitance is proportional to dielectric constant (k), area A, and the inverse of distance d, through the constant $\varepsilon_0$ of $8.854\times10^{-12}$ $\mathrm{F\cdot m^{-1}}$(Equation 1). Thus, the normalized capacitance in thin films, or dielectric constant k will be calculated by using Equation 2. The term "normalized capacitance" is preferred, because this is a study on thin films with morphological variations. In comparison, dielectric constant is an intrinsic property that is free of the impact from such effects. The same measurement is also done on a few known materials (such as paper of similar thickness) for verification and calibration purposes. Five measurements are made for each datapoint.

**Optical Microscopy**

AmScope 40X-2500X LED Trinocular Compound Microscope is used to collect optical micrographs. It has a 3D Two-Layer Mechanical Stage with 18MP USB3.0 Camera. Calibration is performed with OMAX 0.01mm Microscope Camera Calibration Slide (Stage Micrometer).

**Finite Element Analysis (FEA)**

Heat maps were visualized with Matlab simulations, with the following assumed specifications for transient thermal FEA. Thermal conductivities in $W\cdot m^{-1}K^{-1}$: $k_{Air}$=0.025, $k_{Water}$ = 0.6, $k_{NaCl}$ = 6.1, $k_{Surfactant}$ = 1.75 $k_{Water}$ (or 1.2 $k_{Air}$); densities in $kg/m^3$: $\rho_{Air}$ =1.225 (for 15ºC at sea level), $\rho_{Water}$ = 1000, $\rho_{NaCl}$ = 2160, $\rho_{Surfactant}$ = 1330; specific heat capacity in J/(kg·K): $c_{Air}$ =1005, $c_{Water}$ = 4200, $c_{NaCl}$ = 880. The simulation box is a 100-micron square and the NaCl crystal is a 40-micron square at the center. The heat source was set approx. at room temperature in K and the simulation snapshot at 293K. Meshflow was also used to perform FEA of heat map and potential flow in a Android 10 environment. A 2D simulation box of 100 by 100 micron is assumed, along with hypothetical NaCl crystals of cubic shapes and  surfactant coverage of round boundaries.

## TEM

JEOL NEOARM in low beam conditions is used to image SLSA/NaCl samples on continuous carbon films. Although identical slow crystallization may not be possible on TEM grids, a similar concentration and composition ratio is adopted. The final crystal dimensions on TEM grids can be orders of magnitude smaller than the ones on glass slide, but the resultant samples yielded a suitable thickness for TEM, ideal for local structural examination of surfactant mediated salt crystallization. A typical high tension is 80kV, with a spot size 5.

## Semantic Segmentation

Image analysis is performed with Image J. A typical procedure is as follows: (1) convert the optical image type to 8-bit, (2) process the image with the default FFT band filter pass, (3) threshold with slight adjusting to align the highlighted area with the false color enhanced version (generated with Gwyddion).

# Acknowledgement:

This work was supported by Center for Nanophase Materials Sciences (CNMS), which is a US Department of Energy, Office of Science User Facility at Oak Ridge National Laboratory.